\documentclass[letterpaper]{JHEP3}
\pdfoutput=1
\usepackage{graphicx}
\usepackage{amsmath}
\usepackage{dsfont}
\usepackage{subcaption}
\usepackage{amssymb}
\usepackage{bm}
\newcommand{\eq}{\begin{equation}}
\newcommand{\eeq}{\end{equation}}

\title{Axionic charged black branes with arbitrary scalar nonminimal coupling}
\author{Adolfo Cisterna\\
Universidad Central de Chile, Vicerrectoria acad\'emica, Toesca 1783, Santiago, Chile \\
E-mail: \email{adolfo.cisterna-at-ucentral.cl}}

\author{Luis Guajardo\\
Instituto de Matem\'atica y Fisica, Universidad de Talca,
Casilla 747, Talca, Chile.\\
E-mail: \email{luis.guajardo.r-at-gmail.com}}

\author{Mokhtar Hassaine\\
Instituto de Matem\'atica y Fisica, Universidad de Talca,
Casilla 747, Talca, Chile.\\
E-mail: \email{hassaine-at-inst-mat.utalca.cl}}

\abstract{In this paper, we construct four-dimensional charged black
branes  of a nonminimally coupled and self-interacting scalar field.
In addition to the scalar and Maxwell fields, the model involves two
axionic fields homogeneously distributed along the two-dimensional
planar base manifold providing in turn a simple mechanism of
momentum dissipation. Interestingly enough, the horizon of the
solution can be set at two different positions, whose locations depend on the axionic parameter, and in both cases there exists a wide
range of values of the nonminimal coupling parameter yielding
physical acceptable solutions. For one of our solutions, the allowed nonminimal coupling parameters take discrete values and it turns out to be extremal since
its has zero temperature. A complete analysis of the thermodynamical
features of the solutions is also carried out. Finally, thanks to
the mechanism of momentum dissipation, the holographic DC
conductivities of the solutions are computed in term of the black
hole horizon data, and we analyze the effects of the nonminimal
coupling parameter on these conductivities. For example, we notice
that for the non extremal solutions, there always exists a
nonminimal coupling (which is greater than the conformal one in four
dimensions) yielding perfect conductivity in the sense that the
conductivity is infinite. Even more astonishing, the conductivity
matrix for the extremal solutions has a Hall effect-like behavior.}

\begin{document}

%%%%%%%%%%%%%%%%%%%%%%%%
\section{Introduction}
%%%%%%%%%%%%%%%%%%%%%%%%
In the last decades, the ideas underlying the Anti
de-Sitter/Conformal Field Theory (AdS/CFT) correspondence have been
applied to get a better understanding of phenomena that occur in the
condensed matter physics as the quantum Hall effect, the
superconductivity or the superfluidity, see e. g.
\cite{Hartnoll:2007ai,Hartnoll:2008vx}. As a significative example,
we can mention the case of charged hairy black holes with a planar
horizon that may be relevant to describe the behavior of
unconventional superconductors \cite{Hartnoll:2009sz}. In this
scenario, the nonzero condensate behavior of the unconventional
superconductors is mimicked by the existence of a hair at low
temperature that must disappear in the high temperature regime
\cite{Hartnoll:2009sz}. Nevertheless, finding black holes with such
features is an highly nontrivial problem that is rendered even more
difficult by the various no-hair theorems with scalar fields
existing in the current literature, see e. g. \cite{BEK2}.
Fortunately, the precursor works of Refs. \cite{BBM} and \cite{BEK1}
have established that scalar fields nonminimally coupled seem to be
an excellent laboratory in order to escape the standard scalar
no-hair theorems. Indeed, as shown independently in Ref. \cite{BBM}
and Ref. \cite{BEK1}, conformal scalar field nonminimally coupled to
Einstein gravity can support black hole configuration with a
nontrivial scalar field. These black hole solutions have been dubbed
BBMB solution in the current literature. However, the BBMB solution
suffers from some pathology essentially because of the divergence of
the scalar field at the event horizon. This inconvenient makes its
physical interpretation and the problem of its stability a subject
of debate. A way of circumventing this pathology consists in
introducing a cosmological constant whose effect is to precisely
push the singularity behind the horizon, and as a direct
consequence, the scalar field becomes well-defined at the event
horizon \cite{Martinez:2002ru}. It is important to stress that for
the BBMB solution or its extensions with cosmological constant
dubbed as the MTZ solution \cite{Martinez:2002ru, Martinez:2005di},
the parameter $\xi$ that couples nonminimally the scalar field to
the curvature is always the conformal one in four dimensions, namely
$\xi=\frac{1}{6}$, and the horizon topology of the black hole
solutions is either spherical or hyperbolic depending on the sign of
the cosmological constant. There also exist examples of black hole
solutions with the conformal coupling $\xi=\frac{1}{6}$ but with a
potential term that breaks the conformal invariance of the matter
source, see e. g. \cite{Anabalon:2012tu,Ayon-Beato:2015ada}.
Nevertheless, charged black holes with planar horizon topology for a
scalar field nonminimally coupled to Einstein gravity with or
without a cosmological constant are not known\footnote{In higher
dimensions $D>4$, it has been shown that locally AdS black hole
solutions with planar base manifold can emerge for scalar field
nonminimally coupled to Lovelock gravity
\cite{Gaete:2013ixa,Gaete:2013oda,Correa:2013bza}.}. It seems that
extra matter source is needed in order to sustain planar charged
black hole with a nonminimal scalar field. This intuition is based
on the works done in \cite{Bardoux:2012aw,Bardoux:2012tr} where a
planar version of the MTZ solution was rendered possible thanks to
the introduction of two $3-$forms that were originated from two
Kalb-Raimond potentials. Interestingly enough, this construction was
also  extended for arbitrary nonminimal coupling in
\cite{Caldarelli:2013gqa}. Very recently, it has also been shown
that (charged) planar AdS black holes can arise as solutions of
General Relativity with a source composed by a conformal scalar
field together with two axionic fields depending linearly on the
coordinates of the planar base manifold \cite{Cisterna:2018hzf}. The
existence of these planar black holes is mainly due to the presence
of the axionic scalar fields which, in addition of inducing an extra
scale, allow the planar solutions to develop an event horizon. More
precisely, as proved in \cite{Cisterna:2018hzf}, the black hole mass
is related to the parameter associated to the axionic fields, and
hence these axionic (charged) black branes can be interpreted as
extremals in the sense that all their Noetherian charges are fixed
in term term of the axionic intensity parameter. Many other
interesting features are inherent to the presence of axionic fields
for planar solutions. Among others, axionic fields depending
linearly on the coordinates of the planar base manifold provide a
very simple mechanism of momentum dissipation
\cite{Andrade:2013gsa}. From an holographic point of view, this
feature has a certain interest since, as established in Refs.
\cite{Donos:2014cya, Donos:2013eha}, the computation of the DC
conductivities can be uniquely expressed in term of the black hole
horizon data. Mainly because of these results, the study of axionic
black hole configurations in different contexts has considerably
grow up the last time, see e. g. \cite{Ling:2016lis, Ling:2016dck,
Erdmenger:2016wyp, Cisterna:2017jmv, Wang:2018hwg,
Bhatnagar:2017twr, Goldstein:2010aw, Mokhtari:2017vyz}.

In the present, we plan to extend the work done in the conformal
situation \cite{Cisterna:2018hzf} to the case of a four-dimensional
scalar field with an arbitrary nonminimal coupling with two axionic
fields. The model also involves a parameter $b$ that enters in the
scalar potential and in the function that minimally couples the
scalar and the axionic fields, see below (\ref{pot}-\ref{epsilon}).
The range of this extra parameter will be fixed by some reality
conditions as well as by demanding the solutions to have positive
entropy. Asymptotically AdS planar dyonic black hole solutions will
be presented with axionic fields homogenously distributed along the
orthogonal planar coordinates of the base manifold for {\it a
priori} any positive value of the nonminimal coupling parameter
$\xi$. However, the positive entropy condition will considerably
reduce the range of the permissible values of the nonminimal
coupling parameter. As in the conformal case
\cite{Cisterna:2018hzf}, the solutions only contain an integration
constant denoted by $\omega$. Interestingly enough, the location of
the event horizon can be at two different positions, depending on the axionic parameter $\omega$. Moreover, for  $\omega<0$, the range
of permissible values of the nonminimal coupling parameter is
discrete, and the solution is shown to be extremal since its has
zero temperature. Finally, the full DC conductivities associated to
the charged black brane solutions will be computed following the
recipes given in \cite{Donos:2014cya, Donos:2013eha}. One of our
motivations is precisely to identify the impact on the
conductivities  of the nonminimal coupling parameter.

The plan of the paper is organized as follows. In the next section,
we present the model which consists of the Einstein gravity action
with a negative cosmological constant  with a source given by a
self-interacting and nonminimally scalar field coupled to two
axionic fields. In Sec. $3$, the asymptotically AdS planar dyonic
solutions are presented. In Sec. $4$, a detailed analysis of the
thermodynamics properties of the solutions through the Hamiltonian
method is provided allowing to identify correctly the mass of the
dyonic solutions. In the next section, following the perturbative
method presented in \cite{Donos:2014cya, Donos:2013eha}, the full DC
conductivities of the (non) extremal solutions are computed and the
effects of the nonminimal coupling parameter are analyzed. The last
section is devoted to our conclusions.

%%%%%%%%%%%%%%%%%%%%%%%%%%%%%%%%%%%%%%%%%%%%%%%%%%%%%%%%%%%%%%
\section{Model, field equations and black brane solutions \label{solutions}}
%%%%%%%%%%%%%%%%%%%%%%%%%%%%%%%%%%%%%%%%%%%%%%%%%%%%%%%%%%%%%%

We consider a four-dimensional Einstein-Maxwell model with a
negative cosmological constant, a nonminimally coupled and
self-interacting scalar field $\phi$ together with two axionic
fields $\psi_i $ for $i=1, 2$ whose action is given by {\small
\begin{eqnarray}
S = \int d^4x \sqrt{-g} \left( \kappa(R - 2\Lambda) - \frac{1}{4}
F_{\mu\nu}F^{\mu\nu} - \dfrac{1}{2}\partial_{\mu}\phi
\partial^{\mu}\phi - \dfrac{\xi}{2}R\phi^2 - U_b(\phi) -
\dfrac{\varepsilon_b(\phi)}{2} \sum_{i=1}^{2} \partial_{\mu} \psi_i\
\partial^{\mu} \psi_i \right).
\label{action}
\end{eqnarray}}
Here the Maxwell field strength  is
$F_{\mu\nu}=\partial_{\mu}A_{\nu}-\partial_{\nu}A_{\mu}$,   the
nonminimal coupling parameter is denoted by $\xi$ while the
potential $U_b$ and the coupling $\varepsilon_b$ will be given
below. For latter convenience, we shall opt for the following
parametrization of the nonminimal coupling parameter $\xi$
\begin{eqnarray}
\xi=\frac{n-1}{4n},\label{parametrizationxi}
\end{eqnarray}
which maps the region $\xi>0$ to $n\in \mathbb{R} \setminus
[0,1]$\footnote{A simple calculation shows that the range of values
$\xi<0$ is not compatible with the reality condition as defined by
Eq. (\ref{omegaqeqmrelation}) .}. Note that the particular value
$\xi=\frac{1}{4}$ which corresponds to the limit $n\to\infty$ will
be treated separately. The potential $U_b(\phi)$ and the coupling
$\varepsilon_b(\phi)$ associated to the axionic fields depend on a
positive constant denoted by $b$ whose range will be conditioned by
some reality conditions as shown below
\begin{subequations}
\begin{eqnarray}
U_b(\phi) = \dfrac{1-n}{8n\left(1-b\phi^{\frac{2}{n-1}}\right)^4}
\Big(&& 4b^3(n-3)(n-4)\phi^{\frac{2n+4}{n-1}}
-3b^2(n-4)(3n-7)\phi^{\frac{2n+2}{n-1}} \nonumber \\
&&+ 6b(n-3)^2\phi^{\frac{2n}{n-1}} - (n-3)(n-4)\phi^{2} -
6b^{5-n}\phi^{\frac{8}{n-1}} \Big), \label{pot}
\end{eqnarray}
\begin{eqnarray}
\varepsilon_b(\phi) = 1 + \dfrac{n-1}{8n\kappa}\left(
\dfrac{(n-2)\phi^2- (n-3)b\phi^{\frac{2n}{n-1}}-
b^{3-n}\phi^{\frac{4}{n-1}}}{(1-b\phi^{\frac{2}{n-1}})^2} \right).
\label{epsilon}
\end{eqnarray}
\end{subequations}
Before proceeding, we would like to stress that the minimal case
$\xi=0$ or equivalently $n=1$ which has been fully treated in
\cite{Andrade:2013gsa} is clearly excluded from our analysis because
of the form of the expressions  $U_b(\phi) $ and
$\varepsilon_b(\phi)$. Also, we may note that for the conformal
situation $\xi=\frac{1}{6}$ or equivalently $n=3$, the potential
vanishes identically and the axionic minimal coupling
$\varepsilon_b(\phi)=1$, and this situation corresponds to the case
already studied in \cite{Cisterna:2018hzf}.

The field equations obtained by varying the action (\ref{action})
with respect to the metric, the Maxwell field $A_{\mu}$, the scalar
field $\phi$ and the axionic fields $\psi_i$ read respectively
\begin{eqnarray}
&&  \kappa(G_{\mu\nu} + \Lambda g_{\mu\nu}) = \frac{1}{2}
T^{\phi}_{\mu\nu} + \frac{1}{2}T^{\psi}_{\mu\nu} + \frac{1}{2}T^{\text{em}}_{\mu\nu},\nonumber\\
&&  \nabla_{\mu} F^{\mu\nu}  = 0,\\
&& \square \phi = \dfrac{n-1}{4n} R\phi + \dfrac{dU_b}{d\phi} +
\dfrac{1}{2}\dfrac{d\varepsilon_b}{d\phi}
 \sum_{i=1}^{2} g^{\mu\nu}\partial_{\mu}\psi_{i}\partial_{\nu}\psi_{i},\nonumber \\
&& \nabla_{\alpha} \left(\varepsilon_b(\phi)\nabla^{\alpha}\psi_{i}
\right) = 0,\quad i=1,2, \nonumber
\end{eqnarray}
where the different energy-momentum tensors are given by
\begin{eqnarray*}
&& T^{\phi}_{\mu\nu} = \partial_{\mu}\phi \partial_{\nu} \phi -
g_{\mu\nu}\left( \dfrac{1}{2}\partial_{\alpha}
\phi \partial^{\alpha}\phi + U_b(\phi) \right) + \dfrac{n-1}{4n} \left(g_{\mu\nu}\square - \nabla_{\mu}\nabla_{\nu} + G_{\mu\nu}\right) \phi^2, \\
&& T^{\psi}_{\mu\nu} = \varepsilon_b(\phi)\sum_{i=1}^{2}\left(
\partial_{\mu}\psi_{i}\partial_{\nu}\psi_{i} -
\dfrac{1}{2} g_{\mu\nu}\partial_{\alpha}\psi_{i}\partial^{\alpha}\psi_{i} \right),\\
&& T^{\text{em}}_{\mu\nu} =  F_{\mu \sigma}F_{\nu}^{\ \sigma} -
\frac{1}{4}g_{\mu\nu}F_{\rho \sigma}F^{\rho \sigma}.
\end{eqnarray*}

Dyonic black brane solutions of the field equations for $\Lambda=-3$
and with axionic fields homogenously distributed along the
two-dimensional planar base manifold can be found, and are given by
\begin{eqnarray}
\label{sol}
&&ds^2=-f(r)dt^2+\frac{dr^2}{f(r)}+r^2(dx^2+dy^2),\nonumber\\
&& f(r)=\frac{1}{r^2} \left(
r-\dfrac{3\omega}{\sqrt{12\kappa}}\right)\left(r +
\dfrac{\omega}{\sqrt{12\kappa}} \right)^3,\ \ \ \psi_1(x) = \omega x, \ \ \ \psi_2(y)= \omega y\nonumber\\
\\
&&\phi(r)=\Bigg[\frac{\omega}{b\left(\sqrt{12\kappa}r+\omega\right)}\Bigg]^{\frac{n-1}{2}},\qquad
A_{\mu} dx^{\mu}= -\dfrac{q_e}{r} dt + \dfrac{q_m}{2}(x\ dy - y\
dx),\nonumber
\end{eqnarray}
where the intensity of the axionic fields $\omega$ is tied to the
electric and magnetic charges by the relation
\begin{eqnarray}
\omega=\pm \Big(\frac{96\,b^{n-1}\,
n\,\kappa^2(q_e^2+q_m^2)}{n-1-8\,
n\,\kappa\,b^{n-1}}\Big)^{\frac{1}{4}}. \label{omegaqeqmrelation}
\end{eqnarray}

We will now be as exhaustive as possible to enumerate the properties
of these dyonic black brane solutions. For a nonminimal coupling
parameter $\xi$ as parameterized by (\ref{parametrizationxi}), we
have to distinguish between the case $n>1$ corresponding to $\xi\in
]0,\frac{1}{4}[$, and the case $n<0$, i. e. $\xi\in
]\frac{1}{4},\infty[$. For $n>1$, in order to deal with a real
constant $\omega$ as defined by (\ref{omegaqeqmrelation}), the
parameter $b$ of the model must belong to the interval $b\in \,]0,
b_0[$ with
\begin{eqnarray}
b_0=\left(\frac{n-1}{8n\kappa}\right)^{\frac{1}{n-1}}, \label{bo}
\end{eqnarray}
while for $n<0$, one must have $b\in \,]b_0,\infty[$. Also, the
constraint (\ref{omegaqeqmrelation}) allows both sign for the
intensity of the axionic fields $\omega$. This remark has an
interesting consequence concerning the location of the horizon
$r_{+}$ since for $\omega>0$, we have
$r_{+}=\frac{3\omega}{\sqrt{12\kappa}}$ while for $\omega<0$, the
event horizon is located at $r_{+}=-\frac{\omega}{\sqrt{12\kappa}}$.
In this former case, because of the form of the metric function
(\ref{sol}), the temperature of the solution vanishes identically
and the black brane configuration can be interpreted as an extremal
solution. It remains to ensure that the scalar field (\ref{sol}) is
well-defined in the region outside the horizon, namely for $r\in
[r_{+},\infty[$. For $\omega>0$, the scalar field is regular
everywhere in the region $r>0$. On the other hand, for a negative
axionic intensity parameter $\omega<0$, the reality condition on the
scalar field on the region $r\in
[r_{+}=-\frac{\omega}{\sqrt{12\kappa}},\infty[$ restricts the
parameter $n$ to be an odd negative integer $n=1-2k$ with
$k\in\mathds{N}\setminus\left\{0\right\}$ or equivalently the
nonminimal coupling parameters is forced to take the discrete values
given by $\xi=\frac{k}{2(2k-1)}$. All these details are summarized
in the table $1$. It is also interesting to stress that for
$\omega<0$, even if the scalar field vanishes at the horizon, the
expressions of the potential (\ref{pot}) and the coupling
(\ref{epsilon}) remain finite once evaluated on the solution at the
horizon, i. e.
\begin{eqnarray}
U_b(\phi)|_{r_{+}}=\frac{3(n-1)b^{1-n}}{4n},\qquad
\varepsilon_b(\phi)|_{r_{+}}=1-\frac{(n-1)b^{1-n}}{8n\kappa}.
\label{finiteEB}
\end{eqnarray}
This remark will be of importance for the finiteness of the DC
conductivities of the extremal solutions. Finally, we would like to
make a comment concerning the neutral configuration $q_e=q_m=0$. In
this case, the reality condition on $\omega$ given by Eq.
(\ref{omegaqeqmrelation}) is replaced by a constraint on $b$ given
by $b=b_0$ where $b_0$ is defined in (\ref{bo}).

%%%%%%%%%%%%%%%%%%%%%%%%%%%%%%%%%%%%%%%%%%%%%%%%%%%%%%%%%%%%%%%%
\begin{table}%[h!]
\caption{\label{tabla1}Range of the permissible values of  the
nonminimal coupling parameter $\xi$ depending on the sign of the
axionic parameter $\omega$ ensuring a real solution.}
\begin{tabular}{|c|c|c|}
\hline
Sign of $\omega$ & horizon $r_{+}$ & Permissible values of $n$ or equivalently of $\xi$ and range of $b$\\
\hline \hline ~$\omega>0$~ &
~$r_{+}=\frac{3\omega}{\sqrt{12\kappa}}$~ &$n>1$ i. e.
$\xi\in\ ]0,\frac{1}{4}[$\quad $b\in]0,b_0[$\\
[1ex] \hline \hline ~$\omega>0$~ &
~$r_{+}=\frac{3\omega}{\sqrt{12\kappa}}$~ &$n<0$, i. e.
$\xi\in\ ]\frac{1}{4},\infty[,\quad b\in]b_0, \infty[$\\
[1ex] \hline \hline ~$\omega<0$~ &
~$r_{+}=-\frac{\omega}{\sqrt{12\kappa}}$~& $n=1-2k$, i. e.
$\xi=\frac{k}{2(2k-1)}$,\quad $k\in\mathds{N}\setminus\left\{0\right\},\quad b\in]b_0, \infty[$\\
[1ex] \hline
\end{tabular}
\end{table}
%%%%%%%%%%%%%%%%%%%%%%%%%%%%%%%%%%%%%%%%%%%%%%%%%%%%%%%%%%%%%%%%%%

To conclude this section, we briefly report the solution for the
special coupling $\xi=\frac{1}{4}$ which was excluded from the
previous study. In contrast with the other couplings, the black
brane solution can not be charged, and the neutral solution is given
by
\begin{eqnarray}
ds^2 &=& -\left( r^2 - \dfrac{\omega^2}{2\kappa} -
\dfrac{\omega^3}{2\kappa r} \right) dt^2 + \dfrac{dr^2}{\left( r^2 -
\dfrac{\omega^2}{2\kappa}
- \dfrac{\omega^3}{2\kappa r}\right)} + r^2(dx^2 +dy^2),\\
\phi(r)&=& e^{\frac{r}{\omega}},\ \ \ \psi_1(x) = \omega x,\ \ \
\psi_2(y) = \omega y,
\end{eqnarray}
for a potential $U$ and a coupling $\varepsilon$ free of any
couplings that read
\begin{eqnarray}
U(\phi) &=& \dfrac{\phi^2}{4\kappa \ln(\phi)} \left( 2\kappa \ln(\phi)^3 + 6\kappa \ln(\phi)^2 + (3\kappa - 1)\ln(\phi) -2  \right),\\
\varepsilon(\phi) &=& 1 -  \dfrac{\phi^2}{4\kappa}\left(\ln(\phi) +
2 \right).
\end{eqnarray}
One can easily see that for $\omega>0$, the metric solution admits a
single root located at $r_{+}\in ]\frac{\omega^2}{3\kappa},\infty[$.

%%%%%%%%%%%%%%%%%%%%%%%%%%%%%%%%%%%%%%%%%%%%%%%%%%%%%%%
\section{Thermodynamics of the solutions by means of the Hamiltonian method}
%%%%%%%%%%%%%%%%%%%%%%%%%%%%%%%%%%%%%%%%%%%%%%%%%%%%%%%

%Before proceeding with the Hamiltonian method, we would like first
%to clarify the status of the constant $b$ appearing in the potential
%and the coupling (\ref{pot}-\ref{epsilon}). As already mentioned,
%$b$ is a parameter of the theory in the same footing that the
%nonminimal coupling parameter, and consequently there are not
%integration constants that must be varied in the thermodynamics
%analysis. With this respect, the conformal case $\xi=1/6$ or
%equivalently $n=3$ is very peculiar since the potential (\ref{pot})
%as well as the minimal coupling function $\epsilon_b$
%(\ref{epsilon}) do not depend on $b$. Nevertheless, the
%$b-$dependence of the scalar field (\ref{sol}) can be re-expressed
%in terms of the electromagnetic charges and $\omega$ through the
%constraint (\ref{omegaqeqmrelation}) but even if
%$b=b(q_e,q_m,\omega)$ its variation must vanish in accordance with
%the boundary conditions of the scalar field \cite{Cisterna:2018hzf}.

We now  study the thermodynamics properties of the non-extremal
solution (\ref{sol}) with $\omega>0$ whose event horizon is located
at $r_{+}=\frac{3\omega}{\sqrt{12\kappa}}$ with $n\in \mathbb{R}
\setminus [0,1]$ or equivalently $\xi\in]0,\frac{1}{4}[\cup
]\frac{1}{4},\infty[$. In order to achieve this task, we will
proceed using the Euclidean approach where the time is imaginary and
periodic with period $\beta = T^{-1}$. Here, $T$ stands for the
temperature which is fixed by requiring regularity at the horizon.
The temperature of the non-extremal solution (\ref{sol}) reads
\begin{eqnarray}
T = \dfrac{16\omega}{9\pi\sqrt{12\kappa}}. \label{temperature}
\end{eqnarray}
The Euclidean action $I_{\tiny{Euc}}$ is related to the Gibbs free
energy $\mathcal{F}$ by
$$
I_{\tiny{Euc}} = \beta \mathcal{F} = \beta (\mathcal{M} -
T\mathcal{S} - \sum_i \mu_i \mathcal{Q}_i),
$$
where $\mathcal{M}$ is the mass, $\mathcal{S}$ the entropy and
$\mu_i$ are the extra potentials with their corresponding charges
$\mathcal{Q}_i$  \cite{Gibbons:1976ue}.

In order to construct a well-defined Euclidean action, we will
consider a mini superspace where the metric is static and given by
\begin{eqnarray*}
ds^2 = N(r)f(r) d\tau^2 + \dfrac{dr^2}{f(r)} + r^2(dx^2 + dy^2),
\end{eqnarray*}
with $\tau \in [0, \beta]$. The radial coordinate $r$ ranges from
the horizon to infinity, i. e. $r \geq r_{+}$, and the planar
coordinates both are assumed to belong to a compact set, that is
$x\in \Omega_x$ and $y\in \Omega_y$ with $\int dx\,dy=\Omega_x
\Omega_y$. We also assume a specific ansatz for the scalar, axionic
and electromagnetic fields, i. e. $\phi = \phi(r)$, $A_{\mu}dx^{\mu}
= A_{\tau}(r)d\tau + A_{x}(y)dx + A_{y}(x)dy$, and $
\psi_1=\psi_{1}(x)$, and $\psi_{2}= \psi_{2}(y)$. In doing so, the
reduced Euclidean action takes the form
\begin{eqnarray}
I_{\tiny{Euc}}= \int d^4x\ \left(N\mathcal{H} + A_\tau p' \right) +
B,
\end{eqnarray}
where $B$ is a boundary term that will be properly fixed below. Here
$p$ is the conjugate momentum of $A_{\tau}$, $p = -\dfrac{
r^2}{N(r)}A_{\tau}(r)'$, and the reduced Hamiltonian $\mathcal{H}$
is given by
\begin{eqnarray*}
\nonumber \mathcal{H} &=& (2\kappa - \xi \phi(r)^2)(r f'(r) + f(r))
-
\dfrac{1}{2}f(r)r^2(-1+4\xi)\phi'(r)^{2} -r\xi\phi(r) (rf'(r) + 4f(r))\phi'(r)\nonumber\\
&& -2r^2f(r)\xi\phi(r)\phi''(r) + \dfrac{\varepsilon_b(\phi)}{2}
\sum_{i=1}^{2} (\partial_{i} \psi_i)^2 + \dfrac{ (\partial_x A_y -
\partial_y A_x)^2}{2r^2}+ \dfrac{2p^2}{r^2} + r^2U_b(\phi) +
2\kappa r^2 \Lambda.
\end{eqnarray*}
The boundary term $B$ is fixed by requiring that the reduced action
has an extremum, that is $\delta I_{\tiny{Euc}} =0$, yielding to
{\small
\begin{eqnarray}
\label{bdy} \nonumber &&\delta B = \Big[ \left( -A_{\tau}\delta p -
N(r)\left[ (2r\kappa - r\xi\phi(r)^2
r^2\xi\phi(r)\phi'(r)) \delta f - r^2( f'(r)\xi\phi(r) + f(r)\phi'(r) - 2f(r)\xi\phi'(r)) \delta \phi \right. \right. \\
&& \left. \left. + 2r^2 f(r)\xi \phi(r) \delta \phi' \right]
\right)_{r_{+}}^{R} - \int_{r_{+}}^{R}dr\ \varepsilon_b(\phi(r))
\left\lbrace \left[ \int_{\Omega_y}dy\ \partial_{x}\psi_{1} \delta
\psi_1 \right]_{x\in \Omega_x} + \left[ \int_{\Omega_x}dx\
\partial_{y}\psi_{2} \delta \psi_2 \right]_{y\in\Omega_y} \right\rbrace \\
&& \nonumber -  \int_{r_{+}}^{R}dr\ \dfrac{N(r)}{r^2} \left\lbrace
\left[ \int_{\Omega_x}dx\ (\partial_y A_x -
\partial_x A_y)\delta A_x \right]_{y\in\Omega_y} - \left[
\int_{\Omega_y}dy\ (\partial_y A_x - \partial_x A_y)\delta A_y
\right]_{x\in\Omega_x} \right\rbrace\Big]{\beta\Omega_x\Omega_y},
\end{eqnarray}}
where this expression has to be evaluated at the limit $R\to\infty$.
The field equations obtained from varying the Euclidean action
(which do not depend on the boundary term $B$) imply that $N$ is a
constant, and this latter can be chosen without any loss of
generality to be $N(r)=1$. On the other hand,  Gauss law implies
that $p=cst= q_e$. For the axionic fields, we note that their
contribution strongly depends on the integral $\int_{r_{+}}^{r}
\varepsilon_b(\phi(r))\ dr$, which in our case can be computed
yielding to
\begin{equation}
\label{intepsilon} \int_{r_{+}}^{r} \varepsilon_b(\phi(r))dr = r -
\dfrac{(n-1)b^{1-n}}{96\kappa^2 nr}
\left(\omega^{n-1}(\sqrt{12\kappa}r + \omega)^{3-n}- \omega^2
\right) - (r \longleftrightarrow r_{+}).
\end{equation}
The final variation of $\delta B$ at the infinity is finite and
given by
\begin{eqnarray*}
\delta B (\infty) = \beta\Omega_x\Omega_y \left(
\dfrac{4\omega^2}{\sqrt{12\kappa}}  + \eta
\dfrac{(n-3)(n-1)\sqrt{12\kappa} b^{1-n}\omega^2}{48n\kappa^2}
\right),
\end{eqnarray*}
where $\eta$ is defined as
\begin{eqnarray}
\eta=\delta_n^2+\delta_n^{-k},\qquad \mbox{where}\quad
k\in\mathds{N}\setminus\left\{0\right\}. \label{eta}
\end{eqnarray}
The emergence of this extra contribution proportional to $\eta$
which is effective only for $n=2$ and for any nonzero negative
integer can be explained from the fact that the "binomial"
expression appearing between the large brackets in
(\ref{intepsilon}) will contribute with a linear term in the radial
coordinate only for integer values $n\in\mathds{N}$ less or equal to
three\footnote{We may remember that the values $n=0$ and $n=1$ were
excluded from the very beginning, and for $n=3$ the expression
multiplying $\eta$ vanishes identically.}. Working in the grand
canonical ensemble, where $\beta$ and all the potentials are fixed,
the boundary term at the infinity can be integrated as
\begin{eqnarray}
\label{bdyinf} B (\infty) = \beta\Omega_x\Omega_y \left(
\dfrac{4\omega^3}{3\sqrt{12\kappa}}  + \eta
\dfrac{(n-3)(n-1)\sqrt{12\kappa} b^{1-n}\omega^3}{144n\kappa^2}
\right).
\end{eqnarray}

For the contribution at the horizon of the Euclidean action, we
require the following variations
\begin{eqnarray*}
&& \delta f\left|_{r_{+}} = -f'(r_{+})\delta r_{+}, \right. \ \ \
\delta \phi \left|_{r_{+}}= \delta (\phi(r_{+})) - \phi'(r_{+})\delta r_{+} \right. \\
&& \delta p\left|_{r_{+}} = \delta q_e, \right. \ \ \ \delta A_y
\left|_{r_{+}}= \dfrac{x}{2} \delta q_m, \ \ \ \delta A_x\left|_{r_{+}} = -\dfrac{y}{2} \delta q_m \right. \right. \\
&& \delta \psi_1\left|_{r_{+}} = x \delta \omega, \ \ \ \delta
\psi_2\left|_{r_{+}} = y\delta \omega \right. \right.
\end{eqnarray*}
For reason that will become clear, see below Eq. (\ref{entropy}), we
define
\begin{eqnarray}
\tilde{G} = \dfrac{1}{16\pi\kappa- \dfrac{2(n-1)}{n}\pi
\phi(r_+)^2}, \label{G}
\end{eqnarray}
and we finally get
\begin{eqnarray}
\delta B(r_{+})= \delta \left(\dfrac{A_{+}}{4\tilde{G}} \right) +
\beta \left[\Phi_e \delta q_e +  \Phi_m \delta q_m -  \Phi_1 \delta
\omega -  \Phi_2 \delta \omega\right]\Omega_x\Omega_y,
\end{eqnarray}
where $A_{+} = \Omega_x\Omega_y r_{+}^2$ is the horizon area,
$\Phi_e = \dfrac{q_e}{r_{+}}$, $\Phi_m = \dfrac{q_m}{r_{+}}$ and
$\Phi_1$ and $\Phi_2$ correspond to the axionic potentials, defined
as $\omega$ times the horizon term of (\ref{intepsilon}), i.e.,
\begin{eqnarray*}
\Phi_1 = \Phi_2 \equiv \omega \left( r_{+} -
\dfrac{(n-1)b^{1-n}}{96\kappa^2 nr_{+}}
\left(\omega^{n-1}(\sqrt{12\kappa}r_{+} + \omega)^{3-n}- \omega^2
\right) - \eta \dfrac{(n-3)(n-1)\sqrt{12\kappa}
b^{1-n}\omega}{96n\kappa^2} \right).
\end{eqnarray*}
Therefore, the boundary term at the horizon is
\begin{eqnarray}
\label{bdyhor} B(r_{+})= \dfrac{A_{+}}{4\tilde{G}} + \beta
\left[\Phi_e  q_e +  \Phi_m q_m -  \Phi_1 \omega -  \Phi_2
\omega\right]\Omega_x\Omega_y.
\end{eqnarray}
Plugging (\ref{bdyinf}) and (\ref{bdyhor}) we obtain our boundary
term, $B = B(\infty) - B(r_{+})$. The relation between the boundary
term and the Gibbs free energy allows us to identify the mass
$\mathcal{M}$, the entropy $\mathcal{S}$, the electric charge
$\mathcal{Q}_e$, the magnetic charge $\mathcal{Q}_m$ and the axionic
charges $\mathcal{Q}_1$ and $\mathcal{Q}_2$,
\begin{subequations}
\begin{eqnarray}
\label{mass}
\mathcal{M} &=& \Omega_x\Omega_y \left( \dfrac{4\omega^3}{3\sqrt{12\kappa}}  + \eta \dfrac{(n-3)(n-1)\sqrt{12\kappa} b^{1-n}\omega^3}{144n\kappa^2} \right), \\
\label{entropy} \mathcal{S} &=& \dfrac{A_{+}}{4\bar{G}},\\
\nonumber\\
 \mathcal{Q}_e &=& \Omega_x\Omega_y q_e,\ \ \mathcal{Q}_m
= \Omega_x\Omega_y q_m,  \ \ \mathcal{Q}_1 = \mathcal{Q}_2 =
-\Omega_x\Omega_y \omega.
\end{eqnarray}
\end{subequations}

It is reassuring to check that in the conformal case $n=3$, these
thermodynamics quantities reduce to those obtained in
\cite{Cisterna:2018hzf}. On the other hand, even if the first law is
satisfied by means of the constraint (\ref{omegaqeqmrelation}), the
signs of the entropy and the mass must be analyzed carefully.
Firstly, the positivity of the entropy imposes a lower bound for the
parameter $b$; this is because the constant $\tilde{G}$ as defined
by (\ref{G}) must be strictly positive, and this leads to
\begin{eqnarray*}
b >\dfrac{1}{4} \left( \dfrac{n-1}{8n\kappa}
\right)^{\frac{1}{n-1}}=\frac{1}{4}\,b_0,\quad \mbox{for}\,\,
n>1,\qquad \mbox{and}\quad \qquad b<\frac{1}{4}b_0, \quad
\mbox{for}\,\, n<0.
\end{eqnarray*}
Referring to the Table $1$, one can see that the reality condition
imposed by (\ref{omegaqeqmrelation}) for $\omega>0$ make possible to
deal with solutions with positive entropy only if $n>1$ or
equivalently $\xi\in ]0,\frac{1}{4}[$ and the range of the parameter
$b$ must then be reduced to $b\in ]\frac{1}{4}b_0, b_0[$. For the
other values of the nonminimal coupling parameter $n<0$ or
equivalently $\xi>\frac{1}{4}$, the entropy of the solutions turns
out to be always negative because of the reality condition
(\ref{omegaqeqmrelation}). Concerning the mass, it is interesting to
note that the mass is always positive for $\eta=0$ as defined by
(\ref{eta}) or for the conformal case $n=3$. On the other hand, for
$n=2$, the positivity of the mass requires $b>\frac{1}{2}b_0$, and
hence our static solution with $n=2$ can have positive entropy and
mass by demanding the parameter $b$ to belong to the set $b \in
]\frac{1}{2}b_0, b_0[$. Finally, for the remaining values for which
$\eta=1$, namely $n=-1, -2,\cdots$, the solutions will always have
negative mass and entropy.

For the extremal solution corresponding to $\omega<0$ and $\xi$
taking discrete values (cf. Table $1$) with a scalar field vanishing
at the horizon $\phi(r_{+})=0$, it is not safe to consider the
Euclidean approach since the inverse of the temperature is infinite.
Nevertheless, one can compute the entropy of the extremal solution
by means of the Wald's formula \cite{Wald:1993nt} yielding
\begin{eqnarray}
\mathcal {S}_{{\tiny\mbox{extremal}}} = 4\pi\kappa\,\Omega_x\,
\Omega_y\, r_{+}^2. \label{entext}
\end{eqnarray}
It is interesting to note that this result matches with the non
extremal expression (\ref{entropy}) with (\ref{G}) by taking
$\phi(r_{+})=0$. In addition, as it can be seen from Eq.
(\ref{entext}), the entropy of the extremal solution is always
positive and given by the Hawking formula, $S=\frac{{\scriptsize
\mbox{Area}}}{4 G}$, after restoring correctly the value of $\kappa$
in term of the Newton gravitational constant, i. e.
$\kappa=\frac{1}{16\pi G}$. On the other hand, the mass of the
extremal solution is given by
\begin{eqnarray}
\mathcal {M}_{{\tiny\mbox{extremal}}}= \Omega_x\Omega_y \left(
\dfrac{4\omega^3}{3\sqrt{12\kappa}}  +
\dfrac{(n-3)(n-1)\sqrt{12\kappa} b^{1-n}\omega^3}{144n\kappa^2}
\right), \label{massext}
\end{eqnarray}
and it is easy to see that for the permissible discrete values of
the nonminimal coupling parameters $n=1-2k$ with
$k\in\mathds{N}\setminus\left\{0\right\}$ this expression is always
positive for $b>b_0$ (cf Table $1$) even if $\omega<0$.

%%%%%%%%%%%%%%%%%%%%%%%%%%%%%%%%%%%%%%%%%%%%%%%%%%%%%%%%%%%%%%%%
\begin{table}%[h!]
\caption{\label{tabla2}Signs of the entropy and mass of the dyonic
solutions whose reality conditions are fixed by Table $1$ and where
$b_0=\left(\frac{n-1}{8n\kappa}\right)^{\frac{1}{n-1}}$.}
\begin{tabular}{|c|c|c|}
\hline
Sign of $\omega$ &  Permissible values of $\xi$ & Signs of entropy and mass\\
\hline \hline ~$\omega>0$~ & $n>1$\quad\mbox{and}\quad $n\not=2$
\quad i. e.
$\xi\in\ ]0,\frac{1}{4}[\setminus\frac{1}{8}$ & $\mathcal{S}>0$\, \mbox{and}$\,\mathcal{M}>0$,\quad\mbox{for}\quad $b\in]\frac{1}{4}b_0, b_0[$\\
[1ex] \hline \hline ~$\omega>0$~ & $n=2$ \quad i. e.
$\xi=\frac{1}{8}$ & $\mathcal{S}>0$\, \mbox{and}$\,\mathcal{M}>0$,\quad\mbox{for}\quad $b\in]\frac{1}{2}b_0, b_0[$\\
[1ex] \hline \hline ~$\omega>0$~ & $n<0$ \quad i. e.
$\xi\in\ ]\frac{1}{4},\infty[$ & $\mathcal{S}<0$\, \mbox{and}$\,\mathcal{M}<0$,\quad\mbox{for}\quad $b>b_0$\\
[1ex] \hline \hline ~$\omega<0$~ & $n=1-2k$,\quad i. e.
$\xi=\frac{k}{2(2k-1)}$,\quad
$k\in\mathds{N}\setminus\left\{0\right\}$& $\mathcal{S}>0$\,
\mbox{and}$\,\mathcal{M}>0$,\quad\mbox{for}\quad $b>b_0$\\
[1ex] \hline
\end{tabular}
\end{table} \newpage
%%%%%%%%%%%%%%%%%%%%%%%%%%%%%%%%%%%%%%%%%%%%%%%%%%%%%%%%%%%%%%%%%%

%%%%%%%%%%%%%%%%%%%%%%%%%%%%%%%%%%%%%%%%%%%%%%%%%%%%%%%%%%%%%%%%%%%%%%%%%%%%%
\section{Holographic DC Conductivities}\label{conductivity}
%%%%%%%%%%%%%%%%%%%%%%%%%%%%%%%%%%%%%%%%%%%%%%%%%%%%%%%%%%%%%%%%%%%%%%%%%%%%%

In Refs \cite{Donos:2014cya, Donos:2013eha}, it was established that
the full DC conductivities for black holes enjoying a momentum
dissipation can be computed in terms of the black hole horizon data.
The main idea is to construct some conserved currents independent of
the holographic radial coordinate $r$. To that end, we follow the
prescription as described in Refs. \cite{Donos:2014cya,
Donos:2013eha}, and we first turn on the following relevant
perturbations on the black brane solution (\ref{sol})
\begin{eqnarray*}
&&\delta A_{x} = -E_{x}t + a_x(r), \ \ \ \delta A_{y} = -E_{y}t + a_y(r)\\
&&\delta g_{tx} = r^2 h_{tx}(r),\ \ \ \delta g_{rx} = r^2h_{rx}(r), \ \ \ \delta g_{ty} = r^2 h_{ty}(r), \ \ \ \delta g_{ry} = r^2h_{ty}(r)\\
&&\delta \psi_1 = \chi_1(r),\ \ \ \delta \psi_2 = \chi_2(r),
\end{eqnarray*}
where $E_{x}, E_{y}$ are two constants. Consequently, the two
perturbed Maxwell equations read
\begin{eqnarray*}
&&f'a'_{x} + f a''_{x} + q_e h'_{tx} + q_m (f'h_{ry} + fh'_{ry}) = 0,\\
&&f'a'_{y} + f a''_{y} + q_e h'_{ty} - q_m (f'h_{rx} + fh'_{rx}) =
0,
\end{eqnarray*}
which allow to define the following two conserved currents,
\begin{subequations}
\begin{eqnarray}
\label{currentx}
&&J_{x}:= - fa'_{x} - q_eh_{tx} - q_m fh_{ry},\\
\label{currenty} &&J_{y}:= - fa'_{y} - q_eh_{ty} + q_m fh_{rx}.
\end{eqnarray}
\end{subequations}
According to the AdS/CFT correspondence, the holographic DC
conductivities are determined by the conserved currents in the
asymptotic boundary, and since  the expressions obtained above
(\ref{currentx}) and (\ref{currenty}) are independent of the radial
coordinate, we can evaluate them at the horizon $r_{+}$. The next
step is to impose boundary conditions on the linearized
perturbations at the black hole horizon. In order to achieve this
task, it is convenient to use the Eddington-Finkelstein coordinates
$(v,r)$ such that $v = t + \int \frac{dr}{f(r)}$. In this case, the
gauge field will be well-defined by demanding
\begin{eqnarray}
\label{pertgauge} a_{x} = -E_{x}\int \dfrac{dr}{f(r)}, \ \ \ a_{y} =
-E_{y}\int \dfrac{dr}{f(r)},
\end{eqnarray}
while we will require the axionic fields to be constant near the
horizon, and for the metric perturbations we will need to impose
\begin{eqnarray}
h_{rx} = \dfrac{h_{tx}}{f(r)}, \ \ \ h_{ry} = \dfrac{h_{ty}}{f(r)}.
\end{eqnarray}
All the previous conditions can be substituted into the $rx$ and
$ry$ components of the linearized Einstein equations, obtaining a
system of equations for $h_{tx}(r_{+})$ and $h_{ty}(r_{+})$, whose
solutions are given by
\begin{subequations}
\begin{eqnarray}
&&h_{tx} = \left. \dfrac{-E_{y}q_m^3 -
q_m\varepsilon_b(\phi)\omega^2 r^2E_{y} -
  q_e\varepsilon_b(\phi)\omega^2 r^2E_{x}-  q_e^2q_mE_{y}}{16\alpha^2 q_m^4 +
 2 q_m^2 \varepsilon_b(\phi)\omega^2 r^2 + \varepsilon_b(\phi)^2 \omega^4r^4 + 16\alpha^2 q_e^2 q_m^2} \right|_{r=r_{+}}, \\
\label{pertmetric} &&h_{ty} = \left. \dfrac{ q_e^2q_mE_{x} +
 q_m^3E_{x}- q_e\varepsilon_b(\phi)\omega^2r^2
E_{y}+4\alpha q_m\varepsilon_b(\phi)\omega^2 r^2E_{x}}{ q_m^4
+2q_m^2 \varepsilon_b(\phi)\omega^2 r^2 + \varepsilon_b(\phi)^2
\omega^4r^4 + Q_e^2 q_m^2} \right|_{r=r_{+}}.
\end{eqnarray}
\end{subequations}
Since the DC conductivities depend on the location of the horizon,
we will first consider the non extremal case $\omega>0$ where the
the horizon is at $r_{+}=\frac{3\omega}{\sqrt{12\kappa}}$ while the
extremal situation $\omega<0$ and
$r_{+}=-\frac{\omega}{\sqrt{12\kappa}}$ will be treated at the end
of the section.  Hence, the final step is to insert (\ref{pertgauge}
- \ref{pertmetric}) into (\ref{currentx}), (\ref{currenty}) to
obtain the conductivities trough
\begin{eqnarray}
\label{condmat} &&\sigma_{xx} = \dfrac{\partial J_{x}}{\partial
E_{x}} = \dfrac{\tilde{\varepsilon_b}\omega^4 (12 (q_e^2 +
q_m^2)\kappa +9
 \tilde{\varepsilon_b}\omega^4)}{16 q_m^2 (q_e^2 + q_m^2)\kappa^2 + 24 q_m^2 \tilde{\varepsilon_b}\omega^4 \kappa + 9 \tilde{\varepsilon_b}^2 \omega^8}, \nonumber\\
&&\sigma_{xy} = \dfrac{\partial J_{x}}{\partial E_{y}} =
\dfrac{q_eq_m \kappa (64 (q_e^2 + q_m^2)\kappa  +
96\tilde{\varepsilon_b}\omega^4)}{64 q_m^2 (q_e^2 + q_m^2)\kappa^2 + 96 q_m^2 \tilde{\varepsilon_b}\omega^4 \kappa + 36 \tilde{\varepsilon_b}^2 \omega^8},\\
&&\sigma_{yx} = \dfrac{\partial J_{y}}{\partial E_{x}} = -
\sigma_{xy}, \ \ \ \sigma_{yy} = \dfrac{\partial J_{y}}{\partial
E_{y}} =  \sigma_{xx},\nonumber
\end{eqnarray}
where  we have defined
\begin{eqnarray}
\tilde{\varepsilon}_b \equiv \varepsilon_b(\phi(r_{+})) = 1 +
\dfrac{(n-1)(4^{2-n}(3n-5)-1)b^{1-n}}{72n\kappa}.
\end{eqnarray}
The conductivity matrix is antisymmetric in accordance with the
invariance under the SO($2$) symmetry. The purely electric DC
conductivity $\sigma_{DC}$ is given by
\begin{eqnarray}
\label{sigmaDC} \sigma_{DC} && \equiv \sigma_{xx}(q_m=0) = 1+\frac{4
q_e^2\kappa}{3\tilde{\varepsilon}_b \omega^4} \nonumber\\
&& =1 + \dfrac{(n-1)-8n\kappa b^{n-1}}{72n\kappa b^{n-1} +
(n-1)(4^{2-n}(3n-5)-1)},
\end{eqnarray}
where we have explicitly used the constraint
(\ref{omegaqeqmrelation}). We are now in position to analyze the
effects of the parameter $b$ and the nonminimal coupling parameter
$n$ on the electrical DC conductivity. We also recall that the
parameter $b$ is subjected to some reality conditions (cf. Table
$1$) corresponding to the mathematical range of $b$ but also to
physical constraints that ensure the positivity of the mass and
entropy (cf. Table $2$) that we will refer as the physical range of
$b$.

Firstly, it is straightforward to prove that there exists a value of
the nonminimal coupling parameter $n_0\approx 3.4681$ corresponding
to $\xi_0\approx 0.1779$ (\ref{parametrizationxi}) such that for
each $n>n_0$, there exists a precise value of the parameter $b$
denoted by $b_1$ with $\frac{b_0}{4}<b_1<b_0$ (cf. Table $2$) such
that the DC conductivity is strictly positive for $b\in ]b_1, b_0[$
and $\sigma_{DC}$ becomes infinite at $(n, b_1)$. In other words,
this means that for any coupling greater than $n_0$, one can always
chose a parameter $b\in ]b_1, b_0[$ of the theory that yields to
well-defined physical solutions, namely solutions with positive mass
and entropy and having a positive conductivity. Additionally for the
choice $b=b_1$, the model will describe perfect conductivity in the
sense that $\sigma_{DC}\to \infty$ at the point $(n, b_1)$ with
$n>n_0$.

Now, in order to  visualize the influence on the electric DC
conductivity of the parameters $b$ and $n$, we will plot the
graphics of  $\sigma_{DC}$ in function of $b$ and $n$. In Fig.$1$,
we plot the conductivity vs the parameter $b$ for two distinct
values of the nonminimal coupling parameter, namely $n=2<n_0$ and
$n=4>n_0$. The graphic given by Fig.$1$(a) for $n=2$ is in fact
representative of all the cases $1<n\leq n_0$. One can see that for
the physical range of $b$, namely $b\in ]\frac{1}{4}b_0, b_0[$  or
$b\in ]\frac{1}{2}b_0, b_0[$ for $n=2$ (cf. Table $2$), the
conductivity is strictly positive and finite. In fact, the
positivity of the conductivity is always ensured even for the full
mathematical range of $b$, i. e. $b\in ]0, b_0[$. On the other hand,
the graphic represented in Fig.$1$(b) for $n=4$ will be similar for
any value $n>n_0$. In this case, as mentioned before, one can see
the existence of a a vertical asymptote at $b=b_1$ (for $n=4$ we
have $b_1\approx 0.66534$) where the conductivity becomes infinite.
Nevertheless, in contrast with the previous case, the conductivity
is positive only for $b\geq b_1$, and hence part of the physical
range, namely $b\in ]\frac{1}{4}b_0, b_1[$ yields to negative
conductivity.
\begin{figure}[h] \centering
\begin{subfigure}[b]{0.45\textwidth}
\includegraphics[width=\textwidth]{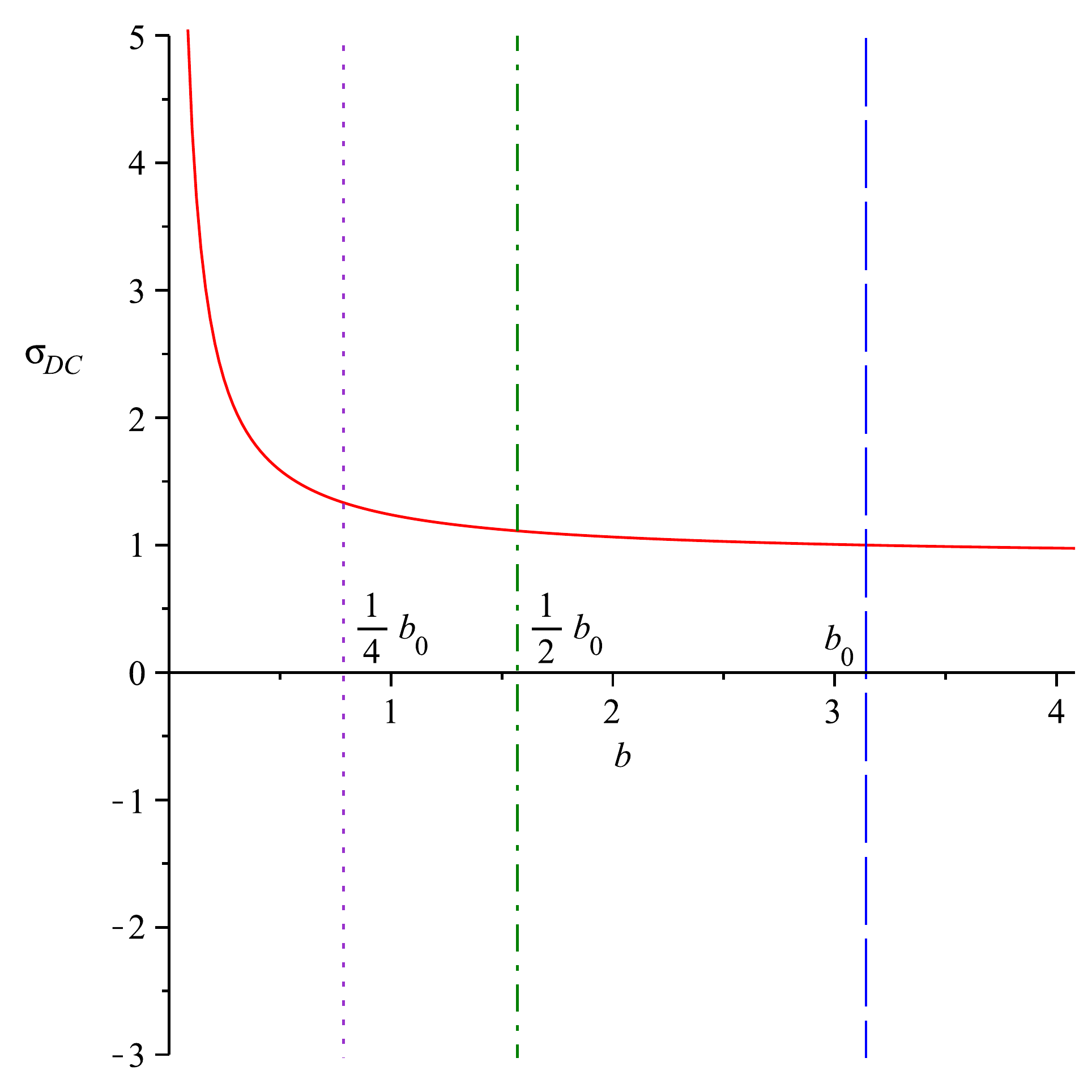}
\caption{$\sigma_{DC}\ \text{vs}\ b\ \text{for}$ $n<n_0$ \quad
\mbox{or equivalently}\quad $ \xi = \frac{1}{8}$}
\end{subfigure}
~ \quad
\begin{subfigure}[b]{0.45\textwidth}
\includegraphics[width=\textwidth]{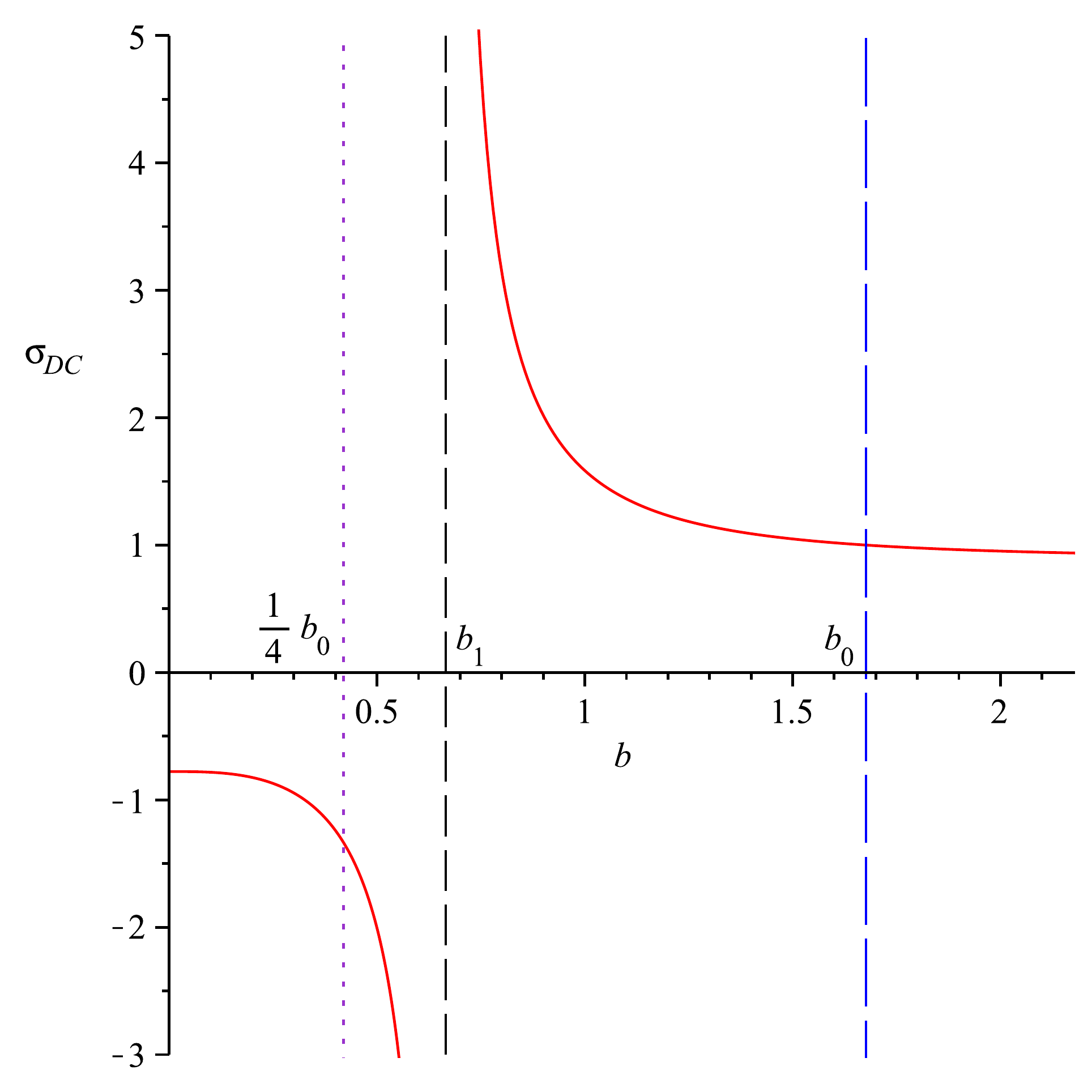}
\caption{$\sigma_{DC}\ \text{vs}\ b\ \text{for}$ $n=4>n_0$ \quad
\mbox{or equivalently}\quad $\xi = \frac{3}{16}$}
\end{subfigure}
\caption{\label{fig1} The conductivity $\sigma_{DC}$ in term of the
parameter $b$ for nonminimal couplings $0<\xi<\frac{1}{4}$.}
\end{figure}
We also include a plot of the non extremal situation for $n<0$ or
equivalently $\xi>\frac{1}{4}$, see Fig.$2$. The mathematical range
of $b$ is located at the right of the blue line while its physical
range is at the left of the dotted line. One can see that for a
mathematically well-defined solution, the electric conductivity is
always positive even if it has negative mass and entropy.
\begin{figure}[h] \centering
\includegraphics[width=0.4\textwidth]{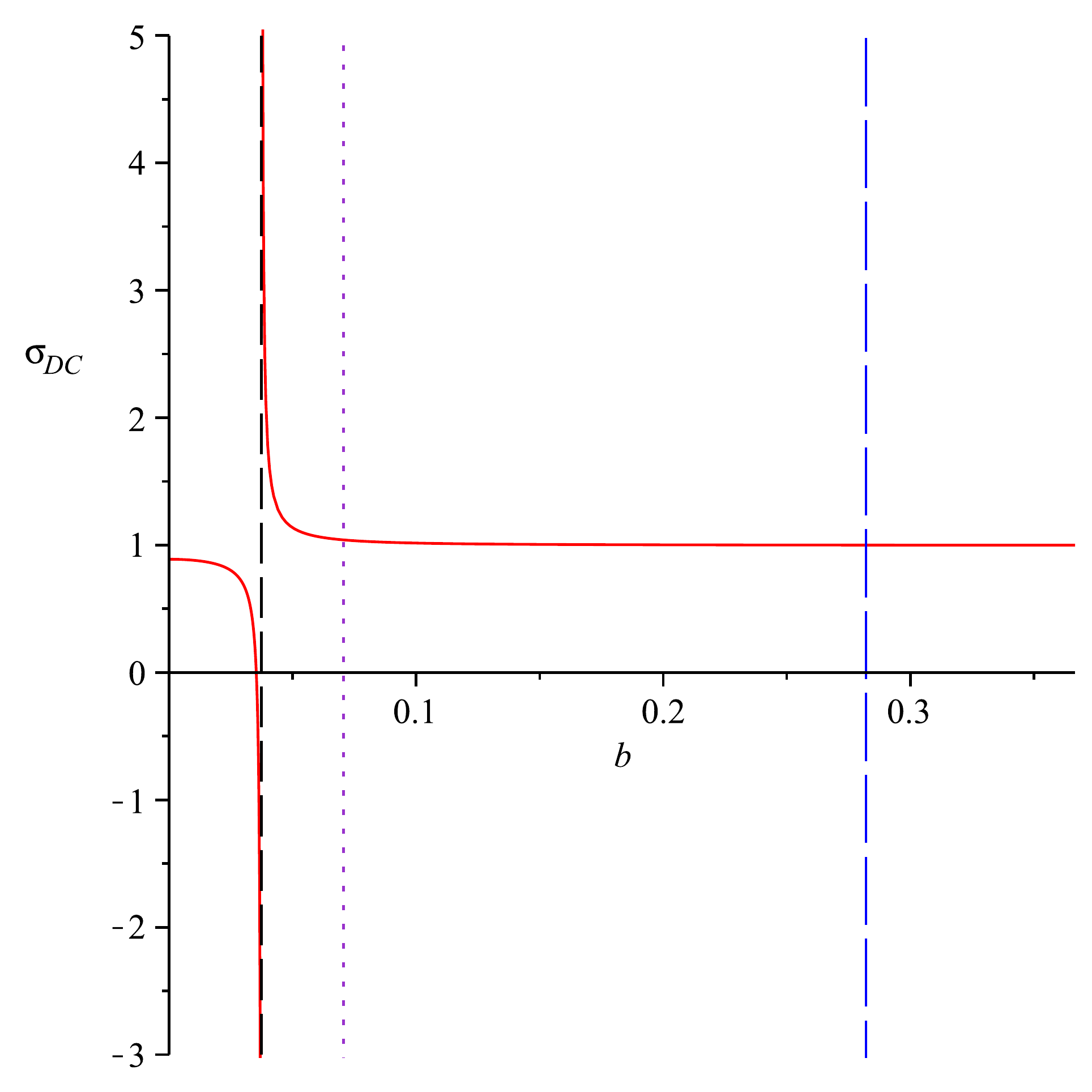}
\caption{\label{fig2} $\sigma_{DC}$ vs $b$ for any nonminimal
couplings $n<0$ or equivalently $\xi>\frac{1}{4}$.}\vspace{1cm}
\end{figure}
We can also appreciate the influence of the nonminimal coupling
parameter on $\sigma_{DC}$ by drawing the graphics of this latter in
function of $n$, see Fig.$3$. In order to achieve this task, we must
be careful with the election of $b$ since its range of permissible
values depends explicitly on $n$, see tables $1$ and $2$ where we
have defined $b_0$ in (\ref{bo}). As explained before, one can see
that for $n\leq n_0$, the electric conductivity is positive while
for $n>n_0$, one has $\sigma_{DC}<0$.
\begin{figure}[h] \centering
\begin{subfigure}[b]{0.45\textwidth}
\includegraphics[width=\textwidth]{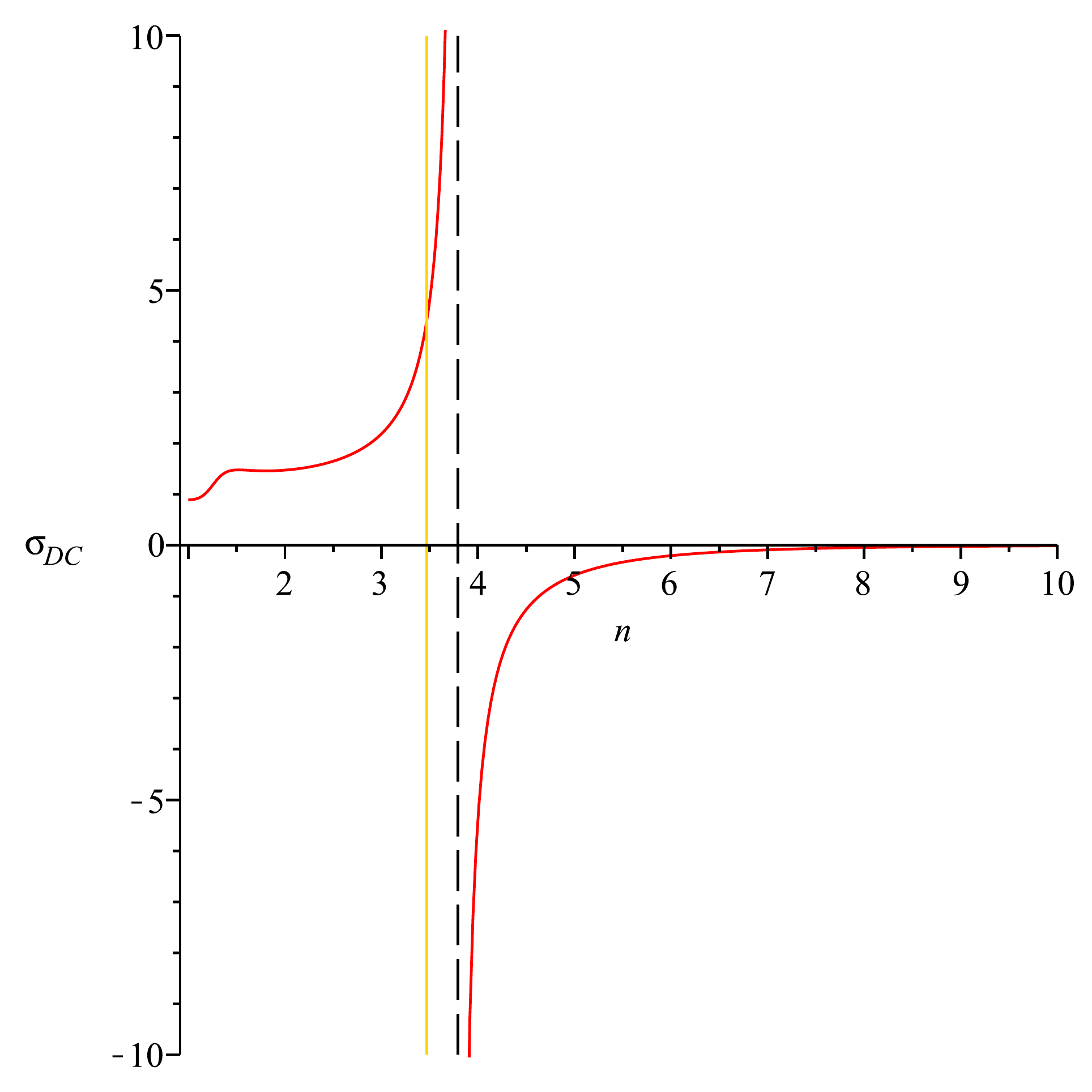}
\caption{$\sigma_{DC}\ \text{vs}\ n\text{ for }b=0.6$}
\end{subfigure}
~ \quad
\begin{subfigure}[b]{0.45\textwidth}
\includegraphics[width=\textwidth]{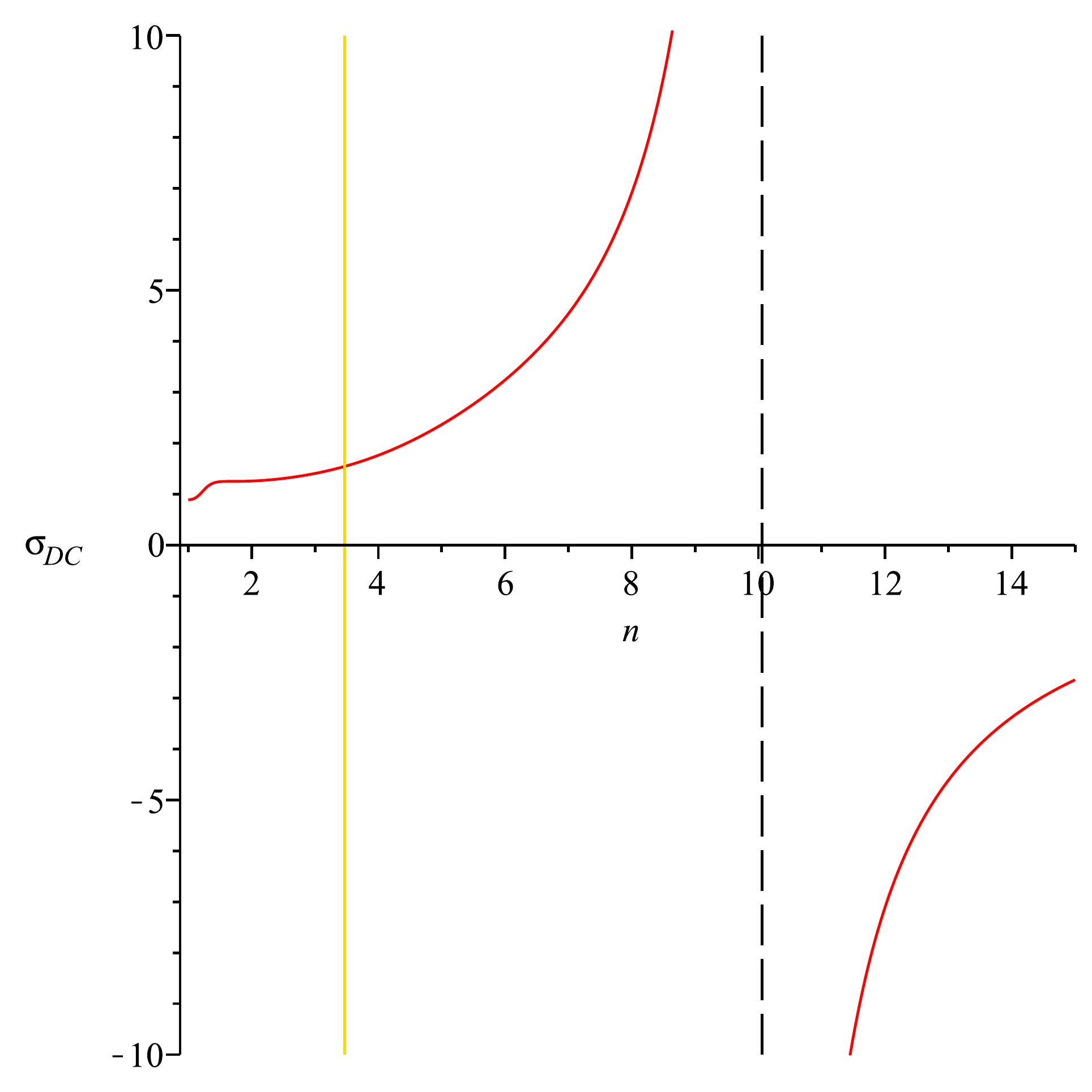}
\caption{$\sigma_{DC}\ \text{vs}\ n\text{ for }b=0.95$}
\end{subfigure}
\caption{\label{fig3} Influence of the nonminimal coupling parameter
$n$ on the electric conductivity.}
\end{figure}
\newpage

The remaining DC conductivities namely the thermoelectric and the
heat conductivities can also be obtained by turning on a more
general time-dependent perturbation\footnote{For simplicity, we will
only consider the DC conductivities along the $x-$coordinate without
magnetic charge.}
\begin{eqnarray*}
\delta A_x=t\left(-E+\zeta a(r)\right)+a_x(r),\,\, \delta
g_{tx}=-t\zeta f(r)+r^2 h_{tx}(r), \,\, \delta g_{rx}=r^2h_{rx},\,\,
\delta\psi_i=\chi(r),
\end{eqnarray*}
where $\zeta$ can be shown to parameterize  a time dependent source
for the heat current. As a consequence, in addition to the conserved
current $J_x$, the following quantity
$$
Q_x=f(r)^2\left(f(r)r^2 h_{tx}\right)^{\prime}-a J_x
$$
is also conserved along the radial coordinate \cite{Donos:2014cya}.
Hence, the thermoelectric conductivities $\alpha$ and $\bar{\alpha}$
as well as the heat conductivity $\bar{\kappa}$ can be computed
yielding
\begin{eqnarray}
\label{alpha} \alpha=\beta\frac{\partial Q_x}{\partial E}=\frac{4\pi
q_e}{\omega^2\tilde{\varepsilon}_b },\qquad
\bar{\alpha}=\beta\frac{\partial J_x}{\partial \zeta}=\frac{4\pi
q_e}{\omega^2\tilde{\varepsilon}_b }\qquad
\bar{\kappa}=\beta\frac{\partial Q_x}{\partial
\zeta}=\frac{64\pi\omega}{3\tilde{\varepsilon}_b
\kappa\sqrt{12\kappa}},
\end{eqnarray}
where $\beta$ as usual is the inverse of the temperature.

Finally, for the extremal solutions, namely $\omega<0$ (cf. Tables
$1$ and $2$), the analysis is completely analogue. This is due to
the fact that, even if the scalar field vanishes at the event
horizon, the minimal coupling function $\tilde{\varepsilon}_b$
remains finite at the horizon (\ref{finiteEB}). As a direct
consequence, the expressions derived previously remain valid with
the difference that in the extremal case, the horizon is located at
$r_{+}=-\frac{\omega}{\sqrt{12\kappa}}$. Interestingly enough, the
conductivity matrix becomes an antisymmetric matrix given by
\begin{eqnarray}
\label{condextremal} \sigma=\left(
\begin{array}{cc}
          0& -\frac{q_m}{q_e}\\
           \frac{q_m}{q_e} & 0
        \end{array}
        \right).
\end{eqnarray}
It is somehow appealing that for the extremal solution, the
conductivity matrix has a  Hall effect-like behavior with a Hall
conductivity that looks like that $B/\rho$ instead of $\rho/B$,
(here $\rho$ is the density of charge and $B$ the orthogonal
magnetic field). Indeed, in the AdS/CFT dictionary,
$(3+1)-$dimensional AdS dyonic black holes are conjectured to be
dual to a $(2+1)$ CFT. In this picture, the electric bulk gauge
field does not have a counterpart in the dual field theory but
instead it fixes the electric charge density $\rho$ to be
proportional to the electric charge of the black hole, i. e.
$\rho\propto q_e$. On the other hand, the magnetic bulk gauge field
is in correspondence with an external magnetic field in the CFT side
with a field strength $B\propto q_m$. Hence, the Hall conductivity
in our case (\ref{condextremal}) is proportional to the ratio
between the magnetic field and the electric charge density, i. e.
$\sigma_{xy}\propto \frac{B}{\rho}$. In order to circumvent this
pathology one could have invoked from the very beginning the
electromagnetic duality symmetry of the dyonic solution which
consists in interchanging the electric with the magnetic charge
namely $q_e\leftrightarrow q_m$, and in this case,one will end with
a Hall conductivity given by $\sigma_{{\tiny\mbox{Hall}}}\propto
\frac{\rho}{B}$.

%%%%%%%%%%%%%%%%%%%%%%%%
\section{Conclusion}
%%%%%%%%%%%%%%%%%%%%%%%%
Here, we have considered a  self-interacting scalar field
nonminimally coupled to the four-dimensional Einstein gravity with a
negative cosmological constant. The matter source is also
supplemented by the Maxwell action with two axionic fields minimally
coupled to the scalar field. Our model is specified from the very
beginning by two parameters that are the nonminimal coupling
parameter denoted $\xi$ or equivalently $n$
(\ref{parametrizationxi}) and the constant $b$ that enters in the
minimal coupling as well as in the potential. For this model, we
have obtained dyonic planar black holes with axionic fields
depending linearly on the coordinates of the planar base manifold.
We have noticed that these charged solutions depend on a unique
integration constant denoted by $\omega$ and the horizon can be
located at two different positions depending on $\omega$. Surprisingly, for $\omega<0$, the temperature of the
solution vanishes identically and hence one can interpret the
solution as an extremal black brane. We have also shown that some
reality conditions (cf. Table $1$) supplemented by the requirement
of having solutions with positive entropy and mass restrict
considerably the permissible values of the nonminimal coupling
parameter and of the parameter $b$, see Table $2$. For a positive
$\omega
>0$, the set of physically acceptable values of the nonminimal
coupling parameter is given by $\xi \in ]0,\frac{1}{4}[$ while for
the extremal solution corresponding to $\omega<0$, only discrete
values of the nonminimal coupling parameter given by
$\xi=\frac{k}{2(2k-1)}$ with
$k\in\mathds{N}\setminus\left\{0\right\}$ yield solutions with
positive mass and entropy. These restrictions on the nonminimal
coupling parameter are to be expected since, even for purely scalar
field nonminimally coupled to Einstein gravity, black hole
configurations have been shown to be ruled out for $\xi<0$ and
$\xi\geq \frac{1}{2}$, see Ref. \cite{Mayo:1996mv}.

In the last part of this work, we have taken advantage of the
momentum dissipation ensured by the axionic fields to compute the
different conductivities by means of the recipes given in Refs.
\cite{Donos:2014cya, Donos:2013eha}. Many interesting results can be
highlighted from the study of the holographic DC conductivities
inherent to these dyonic solutions. For the non extremal solutions,
we have shown that for $n\leq n_0\approx 3.4681$ or equivalently $
\xi\leq \xi_0\approx 0.1779$, the dyonic solutions  always enjoy a
positive conductivity for any mathematically permissible value of
$b$. On the other hand, for $n>n_0$, the positive conductivity
condition restricts the interval of $b$ to be $]b_1,b_0[$ with
$b_1>\frac{1}{4}b_0$. In other words, this means that the physical
solutions (in the sense of having positive mass and entropy) for
$n>n_0$ with $b\in ]\frac{1}{4}b_0, b_1[$ will have a negative
conductivity. Also, we have shown that for $n>n_0$ there always
exist a value of the parameter $b$ denoted by $b_1$ yielding perfect
conductivity in the sense that $\sigma_{DC}(n, b_1)\to \infty$.
Finally, for the extremal solution, we have shown that the diagonal
elements of the conductivity matrix precisely vanish and its
off-diagonal elements are similar to those inherent to the Hall
effect.

An interesting extension of our model will be to consider an
additional $k-$essence term for the axionic part of the action and
to analyze the effects on the conductivities of the nonminimal
coupling parameter $\xi$ conjugated with the $k-$essence parameter,
see Ref. \cite{Cisterna:2017jmv}. In the same lines a natural
generalization of these solutions would be the extension to higher
dimensional scenarios following the lines of \cite{Erices:2017izj}.

Finally, it will be very interesting to explore more deeply some of
the properties of our solutions such as the extremality, the perfect
conductivity or the Hall effect-like behavior of the extremal
solutions. With this respect, in Ref. \cite{Chamblin:1998qm}, it was
shown that the Reissner-Nordstrom at the extremal limit experiences
a sort of Meissner effect in the sense that the magnetic flux lines
are expelled. Hence, a work to be done consists precisely in
investigating the extremal solutions found here can exhibit a kind
of Meissner effect.

%%%%%%%%%%%%%%%%%%%%%%%%%%%%%%%%%%%%%%%%%%%%%%%%%%%%%%%%%%%%%%%%%%%%%%%%%%%%%%%%%%%%%%%%%%%%%%%%%%%%%%%%%
\begin{acknowledgments}
This work has been partially supported by grant FONDECYT 11170274 (A.C). We would
like to thank especially Moises Bravo-Gaete for stimulating
discussions and nice comments to improve the draft. One of us, M. H.
would like to dedicate this work to the memory of his late friend
and professor Christian Duval.
\end{acknowledgments}
%%%%%%%%%%%%%%%%%%%%%%%%%%%%%%%%%%%%%%%%%%%%%%%%%%%%%%%%%%%%%%%%%%%%%%%%%%%%%%%%%%%%%%%%%%%%%%%%%%%%%

%%%%%%%%%%%%%%%%%%%%%%%%%%%

\end{document}